\pdfoutput=1

\documentclass[pdflatex,sn-vancouver,Numbered]{sn-jnl}

\usepackage{graphicx}%
\usepackage{fancyvrb}
\usepackage{multirow}%
\usepackage{amsmath,amssymb,amsfonts}%
\usepackage{amsthm}%
\usepackage{mathrsfs}%
\usepackage[title]{appendix}%
\usepackage[dvipsnames]{xcolor}
\usepackage{textcomp}%
\usepackage{manyfoot}%
\usepackage{booktabs}%
\usepackage{algorithm}%
\usepackage{algorithmicx}%
\usepackage{algpseudocode}%
\usepackage{listings}%
\usepackage{comment}
\usepackage{rotating}
\usepackage{xcolor}
\usepackage{graphicx}
\usepackage{subcaption}
\usepackage{booktabs}
\usepackage{multirow}
\usepackage{siunitx}

\begin{document}

\title[LLaMA Vs ChatGPT]{Can Large Language Models Understand Molecules?}

\author*[1]{\fnm{Shaghayegh} \sur{Sadeghi}}\email{sadeghi3@uwindsor.ca}
\author[1]{\fnm{Alan} \sur{Bui}}
\author[1]{\fnm{Ali} \sur{Forooghi}}
\author[1]{\fnm{Jianguo} \sur{Lu}}
\author[1]{\fnm{Alioune} \sur{Ngom}}

\affil*[1]{\orgdiv{School of Computer Science}, \orgname{Univeristy of Windsor}, \orgaddress{\street{ Sunset Ave}, \city{Windsor}, \postcode{N9B 3P4}, \state{Ontario}, \country{Canada}}}


\abstract{\textbf{Purpose:} Large Language Models (LLMs) like GPT (Generative Pre-trained Transformer) from OpenAI and LLaMA (Large Language Model Meta AI) from Meta AI are increasingly recognized for their potential in the field of cheminformatics, particularly in understanding Simplified Molecular Input Line Entry System (SMILES), a standard method for representing chemical structures. These LLMs also have the ability to decode SMILES strings into vector representations.

\textbf{Method:} We investigate the performance of GPT and LLaMA compared to pre-trained models on SMILES in embedding SMILES strings on downstream tasks, focusing on two key applications: molecular property prediction and drug-drug interaction prediction.

\textbf{Results:} We find that SMILES embeddings generated using LLaMA outperform those from GPT in both molecular property and DDI prediction tasks. Notably, LLaMA-based SMILES embeddings show results comparable to pre-trained models on SMILES in molecular prediction tasks and outperform the pre-trained models for the DDI prediction tasks.

\textbf{Conclusion:} The performance of LLMs in generating SMILES embeddings shows great potential for further investigation of these models for molecular embedding. We hope our study bridges the gap between LLMs and molecular embedding, motivating additional research into the potential of LLMs in the molecular representation field. GitHub: \href{https://github.com/sshaghayeghs/LLaMA-VS-GPT}{https://github.com/sshaghayeghs/LLaMA-VS-GPT}}

\keywords{Large Language Models, LLaMA, GPT, SMILES Embedding}



\maketitle

\section{Introduction}\label{sec1}
Molecule embedding is an important task in drug discovery \cite{li2020learn, lv2021mol2context}, and finds wide applications in related tasks such as molecular property prediction \cite{liu2023molrope,ross2022large,zhang2021mg,chithrananda2020chemberta}, drug-target interaction (DTI) prediction \cite{zhou2021multidti,thafar2022affinity2vec,jin2021embeddti} and drug-drug interaction (DDI) prediction \cite{purkayastha2019drug,han2022smilegnn}.

Molecule embedding techniques learn the features either from the molecular graphs that encode the connectivity information of a molecule structure or from the line annotations of their structures, such as the popular SMILES (simplified molecular-input line-entry system) representation \cite{ross2022large}. 

Molecule embedding via SMILES strings evolve and synchronize with the advances in language modelling \cite{devlin2018bert,vaswani2017attention}, starting with static word embedding \cite{jaeger2018mol2vec}, to contextualized pre-trained models \cite{wang2019smiles, fabian2020molecular, ross2022large}. These embedding techniques aim to capture relevant structural and chemical information in a compact numerical representation \cite{koge2021embedding}. The fundamental hypothesis asserts that structurally similar molecules behave in similar ways. This enables machine learning algorithms to process and analyze molecular structures for property prediction and drug discovery tasks.

With the breakthroughs made in LLMs, one prominent question is whether LLMs can understand molecules and make inferences on molecule data? More specifically, can LLMs produce high quality semantic representations? \cite{guo2023indeed} made a preliminary study by evaluating several chemical inference tasks using LLMs. Their study has been limited to utilizing and evaluating LLMs performance in answering SMILES-related queries. We move further by exploring the ability of these models to effectively embed SMILES has yet to be fully explored, maybe partially due to cost of API calls. Our conclusions are: 
\begin{itemize}
\item[1)] LLMs do outperform traditional methods .
\item[2)] The performance is task dependent, sometimes data dependent. 
\item[3)] Newer versions of LLMs do improve over older versions, even though they are trained on more generic tasks. 
\item[4)] We observe that embeddings from LLaMA overall outperform GPT embeddings. 
\item[5)] Another interesting observation of our research is that LLaMA and LLaMa2 are very close regarding embedding performance.
\end{itemize}

\section{Related Work}
For accurate prediction of chemical properties using machine learning, leveraging molecule embeddings as input feature vectors is crucial \cite{goh2017smiles2vec}. 
Early molecular embedding methods such as Morgan FingerPrint (FP) \cite{morgan1965generation} encode the structural information of a molecule into a fixed-length binary or integer vector with the knowledge of chemistry. 

However, for a more generalized embedding, numerous studies have explored methods to embed molecular structures. While some studies focus on the graph representation of the molecular structure to encode the important topology information directly \cite{duvenaud2015convolutional,wang2022molecular,zang2023hierarchical}, many choose the string representation of molecules (SMILES) due to rapid advancements in natural language processing (NLP). Initial efforts in this domain utilized foundational NLP architectures like auto-encoders \cite{xu2017seq2seq} and recurrent neural networks (RNN) to generate embeddings \cite{goh2017smiles2vec}. However, the scarcity of labelled data has shifted focus towards methods that can be pre-trained on unlabeled data, such as Mol2Vec and SPVec \cite{jaeger2018mol2vec, zhang2020spvec}.

With the increasing prominence of transformer models in natural language analysis—where they are pre-trained on extensive unsupervised data and then fine-tuned for specific tasks like classification—transformer-based models have become increasingly relevant in the SMILES language domain. For instance, SMILES-BERT \cite{wang2019smiles} has inspired numerous studies to adapt the transformers framework. These studies try to modify this framework to improve their performance on SMILES strings by adapting RoBERTa (Robustly optimized BERT approach) instead of the BERT model \citep{chithrananda2020chemberta} or develop domain-specific self-supervised pre-training tasks \cite{fabian2020molecular}, or integrate the local message passing mechanism of graph neural networks (GNNs) into BERT to enhance learning from molecular graphs \cite{zhang2021mg}. Additionally, MolFormer \cite{ross2022large} introduces a novel approach by combining molecular language with transformer encoder models, incorporating rotary positional embeddings (RoPE) from RoFormer, to produce more effective molecular embeddings \cite{ross2022large, su2021roformer}.

However, pre-training these models on millions of molecules requires substantial hardware resources. For example, MolFormer necessitates up to 16 V100 graphics processing units (GPUs) \cite{ross2022large}. Consequently, it is computationally more feasible to use pre-trained large language models (LLMs), such as GPT \cite{radford2019language} and LLaMA \cite{touvron2023llama, touvron2023llama2}, for generating embeddings. These models have already been trained on vast amounts of data, making them readily available for processing SMILES strings to obtain molecular embeddings without extensive hardware.

Up to our current knowledge, the application of GPT and LLaMA in chemistry has primarily been limited to utilizing and evaluating its performance in answering queries. Further exploration and implementation of LLMs for more advanced tasks within chemistry are yet to be thoroughly documented. For example, to examine how well LLMs understand chemistry, Guo et al. \cite{guo2023indeed} used LLMs to assess the performance of these models on practical chemistry tasks only using queries. Their results demonstrate that GPT models are comparable with classical machine learning models when applied to chemical problems that can be transformed into classification or ranking tasks such as property prediction. 
However, they stop evaluating the LLM's ability to answer prompts and do not evaluate the embedding power of LLMs. Hence, inspired by many language-based methods that tried to extract molecular embedding, our study represents a pioneering effort, being the first to rigorously assess the capabilities of LLMs like GPT and LLaMA in using LLMs embedding for chemistry tasks.

\section{LLMs}
\begin{figure}[ht]
    \centering
    \begin{subfigure}[b]{0.4\textwidth} 
        \centering
        \includegraphics[width=\textwidth]{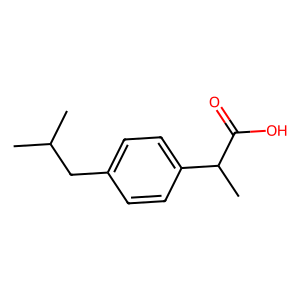} 
        \caption{Kekule Diagram}
    \end{subfigure}%
    \hfill
    \begin{subfigure}[b]{0.4\textwidth} 
        \centering
        \includegraphics[width=\textwidth]{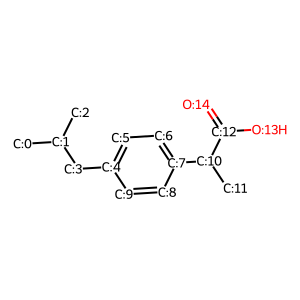} 
        \caption{Kekule Diagram With Properties}
    \end{subfigure}%
    \hfill
    \begin{subfigure}[b]{0.4\textwidth} 
        \centering
        \includegraphics[width=\textwidth]{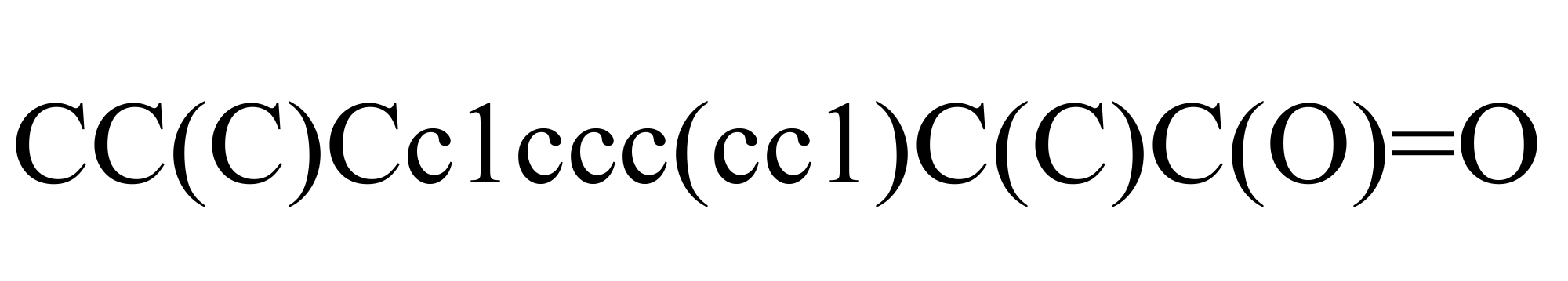} 
        \caption{SMILES String}
    \end{subfigure}%
    \hfill
    \begin{subfigure}[b]{0.4\textwidth} 
        \centering
        \includegraphics[width=\textwidth]{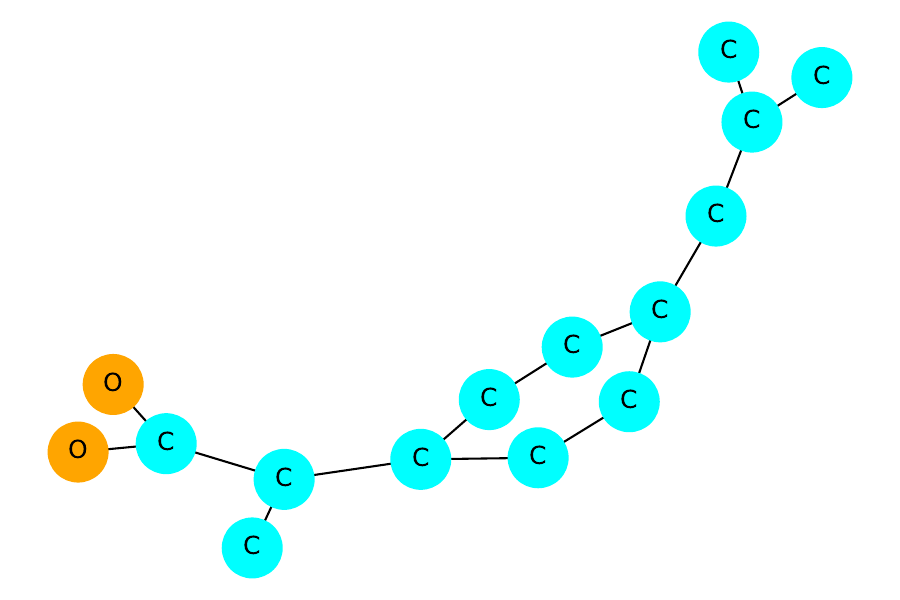} 
        \caption{Molecular Graph}
    \end{subfigure}
    \caption{Drug Chemical Representations.}
    \label{fig:1}
\end{figure}
LLMs, exemplified by architectures like BERT \cite{devlin2018bert}, GPT \cite{radford2019language}, LLaMA \cite{touvron2023llama}, and LLaMA2 \cite{touvron2023llama2} excel at understanding context within sentences and generating coherent text. They leverage attention mechanisms and vast training data to capture contextual information, making them versatile for text generation, translation, and sentiment analysis tasks. While Word2Vec enhances word-level semantics, language models provide a deeper understanding of context and facilitate more comprehensive language understanding and generation.
Pre-trained models from LLMs can transform text into dense, high-dimensional vectors, which capture contextual information and meaning. Using pre-trained LLMs offers an edge as they transfer knowledge from their vast training data, enabling the extraction of context-sensitive representations without requiring extensive task-specific data or feature engineering \cite{hassani2023role}.

This work focuses on obtaining the embeddings of SMILES strings from GPT and LLaMA models to find the model that achieves the best performance. OpenAI \cite{OpenAI} present many GPT-based embeddings including: \textit{'text-embedding-ada-002'}, \textit{'text-embedding-3-small'}, \textit{'text-embedding-3-large'}. Our research used the most recent embedding model, \textit{text-small-3-embeddings}. This model is acclaimed for being the best among available embedding models and the most affordable method available by OpenAI. Text-small-3-embeddings employs the \textit{'cl100k-base'} token calculator to generate embeddings, resulting in a 1536-dimensional vector representation. We input SMILES strings into this model, allowing GPT to create embeddings for each string. These embeddings serve as the feature vector for our classification tasks.

In parallel, we leveraged the capabilities of LLaMA \cite{touvron2023llama} and its advanced variant, LLaMA2 \cite{touvron2023llama2}. These models, ranging from only 7 to 65 billion parameters, are built on the Transformers architecture. LLaMA2, an enhancement of LLaMA, benefits from training on an expanded publicly available data set. Its pre-training corpus grew by 40\%, and its context length doubled to 4096 tokens. LLaMa models employ a decoder-only Transformer architecture with causal multi-headed attention in each layer. Drawing architectural inspiration from prominent language models like GPT-3 and PaLM (Pathways Language Model) \cite{chowdhery2023palm}, they incorporate features such as pre-normalization, RMSNorm, SwiGLU activation functions, and rotary positional embeddings (RoPE) \cite{su2021roformer} in every transformer layer.

The training dataset of LLaMA \cite{touvron2023llama, zhao2023survey} predominantly comprises webpages, accounting for over 80\%  of its content. This is supplemented by various sources, including 6.5\% code-centric data from GitHub and StackExchange, 4.5\% literary content from books, and 2.5\% scientific material primarily sourced from arXiv. 

In contrast, GPT \cite{brown2020language, zhao2023survey} was developed using a comprehensive and mixed dataset. This dataset includes diverse sources like CommonCrawl, WebText2, two different book collections (Books1 and Books2), and Wikipedia. 
 
SMILES is utilized as a "chemical language" that encodes the structural elements of a chemical graph—including atoms, bonds, and rings—into a brief textual format. This is achieved through a systematic, depth-first tree traversal of the chemical structure. The method uses alphanumeric characters to represent atoms (such as C, S, Br) and symbols such as '-', '$=$', and '$\#$' to indicate different types of chemical bonds. For instance, the SMILES notation for Ibuprofen is \textit{CC(C)Cc1ccc(cc1)C(C)C(O)=O} (Figure \ref{fig:1}). 

Table \ref{tab:1} compares how each model tokenizes SMILES strings. ChemBERTa, explicitly designed for molecular embeddings, tokenizes SMILES using the Byte-Pair Encoder (BPE) strategy. Meanwhile, MolFormer-XL employs a SMILES-specific regular expression method, as described by Schwaller et al. \cite{schwaller2019molecular}, using an atom-wise tokenization strategy with the regular expression pattern that is formatted as follows and is able to differentiate between atom characters and symbols for chemical bonds:

\begin{Verbatim}[fontsize=\scriptsize]
(\[[^\]]+]|Br?|Cl?|N|O|S|P|F|I|b|c|n|o|s|p|\(|\)|\.|=|#||\+|
\\\\\/|:|~|@|\?|>|\*|\$|\%[0-9]{2}|[0-9])
\end{Verbatim}

However, LLaMA, as a general-purpose model, employs a different tokenization approach. Its tokenizer is based on SentencePiece Byte-Pair Encoding (BPE). This tokenizer processes the input string character by character, searching for the largest known subword units it can match based on its training. Consequently, as it can be seen in Table \ref{tab:1}, it treats 'CS' from the 'CCS(=O)(=O)CCBr' string as a single token, possibly interpreting it as an abbreviation in natural language. However, 'C' and 'S' should be considered as separate tokens, since each represents a distinct atom.

Table \ref{tab:2} compares molecular embedding in terms of the number of layers, parameters and their speed in generating a SMILES embedding. Compared with Morgan FP, language models are extremely slow. However, GPT performs the fastest among the language models, while LLaMA models are the slowest. There is also a relation between the number of layers and the speed of embedding generation. Although GPT remains an exception.


\begin{table*}
\centering
\tiny
\caption{Comparison of Tokenizers for Molecular SMILES String }
\begin{tabular}{ lll } 
 \toprule
 \textbf{Model}& Tokenization Strategy& Example Tokenization of 'CCS(=O)(=O)CCBr' \\\midrule
 \textbf{BERT Tokenizer} &Subword-based tokenization& ['CC', '\#\#S', '(', '=', 'O', ')', '(', '=', 'O', ')', 'CC', '\#\#B', '\#\#r'] \\ 
 \textbf{GPT Tokenizer} &cl100k-base& ['CC', 'S', '(', '=', 'O', ')(', '=', 'O', ')', 'CC', 'Br']\\
 \textbf{LLaMA2 Tokenizer} &SentencePiece Byte-Pair Encoding-based& ['\textunderscore C', , 'CS', '(', '=', 'O', ')(', '=', 'O', ')', 'CC', 'Br'] \\ 
 \textbf{ChemBERTa Tokenizer} &Byte-Pair Encoding-based& ['C', 'C', 'S', '(', '=', 'O', ')', '(', '=', 'O', ')', 'C', 'C', 'B', 'r'] \\ 
 \textbf{MolFormer-XL Tokenizer} &SMILE Regex& ['C', 'C', 'S', '(', '=', 'O', ')', '(', '=', 'O', ')', 'C', 'C', 'Br'] \\ 
 \bottomrule
\end{tabular}
\label{tab:1}
\end{table*}

\begin{table*}
\centering
\tiny
\caption{Comparison of Embedding Models Used in This Study. $^*$ Speed of Generating Embedding. Speed is Dependent on the Machine.}
\begin{tabular}{ lcccc } 
 \toprule
 \textbf{Model} &\textbf{Dim. Size}& \# Layers & \# Parameters& Speed$^*$\\\midrule
 \textbf{Morgan FP (Radius=2)} &1024& Not applicable & Not applicable& 0.0015 s\\\midrule
\textbf{BERT} &768& 12 & 110 M& 2.9777 s\\
\textbf{ChamBERTa}&384 & 3 & 3 M& 4.8544 s\\
\textbf{MolFormer} &768& 12 & 44 M& 20.9644 s\\ 
\textbf{GPT} &1536& 96 & 175 B& 0.2597 s\\
\textbf{LLaMA}&4095& 32 & 7 B&50.8919 s\\
\textbf{LLaMA2} &4095& 32 & 7 B& 51.6308 s\\
 \bottomrule
\end{tabular}
\label{tab:2}
\end{table*}

\section{Experiments}

Our study aims to generate molecular representation via LLMs and then evaluate the representation on various downstream tasks. To demonstrate the effectiveness of LLMs' molecular representations, we benchmarked their performance on numerous challenging classification and regression tasks from MoleculeNet \cite{wu2018moleculenet} as well as link prediction from BioSnap \cite{biosnapnets} and DrugBank \cite{wishart2018drugbank}. The objective of link prediction in this research is to map the drugs as nodes and their interactions as edges and identify whether there is a missing edge between two drug nodes.

\subsection{Experimental Setup}
We experimented with seven models, each evaluated by six classifications, three regression and two link prediction tasks. To generate embeddings from LLaMAs, BERT, ChemBERTa, and MolFormer models, we first download and load the model weights using the Transformers library and then generate the embeddings. For LLaMA weights, we download the weights provided by Meta for LLaMAs and then convert them into PyTorch format. We extract embeddings from the last layer of the LLMs, following the practice in \cite{reimers2019sentence}. Pooling strategies can impact performance, and we explored a variety of combinations. The overall result remains the same. Hence, for the sake of simplicity, we use only the last layer. For GPT embeddings, we choose the recent model, \textit{text-small-3-embeddings}.

To generate LLaMA and LLaMA2 embeddings, we employed four NVIDIA A2 GPUs to load the 7 billion parameter version of LLaMAs. In this configuration, the average speed of generating embeddings is one molecule per second. In our experiments, we generated embeddings for over 65,000 molecules.

Following MoleculeNet \cite{wu2018moleculenet}, for classification tasks, we partition the datasets into 5-stratified folds to ensure robust benchmarking. This approach ensures that each fold maintains the same proportion of observations for each target class as in the complete dataset. We employ a logistic regression model from scikit-learn, equipped with the following default parameters: L2 regularization, 'lbfgs' for optimization, and maximum 100 iterations allowed for the solvers to converge. The reported performance metrics are the mean and standard deviation of the F1-score and AUROC, calculated across the five folds.

For regression tasks, we implement 5-fold cross-validation to assess model performance. We employ a Ridge regression model which is a linear regression model with l2 regularization. From scikit-learn with the following default parameters:  tolerance of 0.001 for the optimization and a auto solver to automatically chooses the most appropriate solver method based on the data type. The metrics reported are the mean and standard deviation of the RMSE and the R$^2$, calculated across the five folds.

Following MIRACLE \cite{wang2021multi}, a state-of-the-art method in DDI, for link prediction, we split all interaction samples from the DrugBank and BioSnap datasets into training and test sets using a 4:1 ratio. We further select 1/4 of the training dataset as a validation set. The reported results are the mean and standard deviation of AUROC and AUPR across 10 different runs of the GCN model.
We set each parameter learning rate using an exponentially decaying schedule with an initial learning rate of 0.0002 and a multiplicative factor of 0.96.
For the proposed model’s hyperparameters, we set the dimension of the hidden state of drugs as 256 and 3 layers for the GCN encoder. To further regularise the model, dropout with p = 0.3 is applied to every intermediate layer’s output. We use Pytorch-geometric \cite{fey2019fast} for GCN. GCN Model is trained using Adam optimizer.

\subsection{Benchmarking Data Sets}
For classification and regression tasks, we use datasets from MoleculeNet \cite{wu2018moleculenet}, which is a collection of diverse datasets that cover a range of tasks, such as identifying properties like toxicity, bioactivity, and whether a molecule is an inhibitor. MoleculeNet is a widely used benchmark dataset in the field of computational chemistry and drug discovery and it is designed to evaluate and compare the performance of various machine learning models and algorithms on tasks related to molecular property prediction, compound screening, and other cheminformatics tasks \cite{chithrananda2020chemberta,li2021mol,zhang2021mg,ross2022large,liu2023molrope,zang2023hierarchical,guo2023indeed}.

For the link prediction task, however, we utilize two DDI networks: BioSnap \cite{biosnapnets} and DrugBank \cite{wishart2018drugbank}. These datasets represent interactions among FDA-approved drugs as a biological network, with drugs as nodes and interactions as edges.

We extracted the SMILES strings of drugs in the DrugBank database. It should be noted that we conduct data removal because of some improper drug SMILES strings in Drugbank, which can not be converted into molecular graphs, as determined by the RDKit library. The errors include so-old storage format of SMILES strings, wrong characters, etc. Through these curation efforts, we have fortified the quality and coherence of our DDI network, ensuring its suitability for comprehensive analysis and interpretation.

For the BioSnap dataset, 1320 drugs have SMILES strings, while the DrugBank dataset has 1690 drugs with SMILES strings. Hence, the number of edges for BioSnap and DrugBank reduced to 41577 and 190609, respectively.

\subsection{Performance Analysis}
\subsubsection{Results on Classification Tasks}
\begin{figure}[H]
    \centering
    \begin{subfigure}[b]{\textwidth} 
        \centering
        \includegraphics[width=\textwidth]{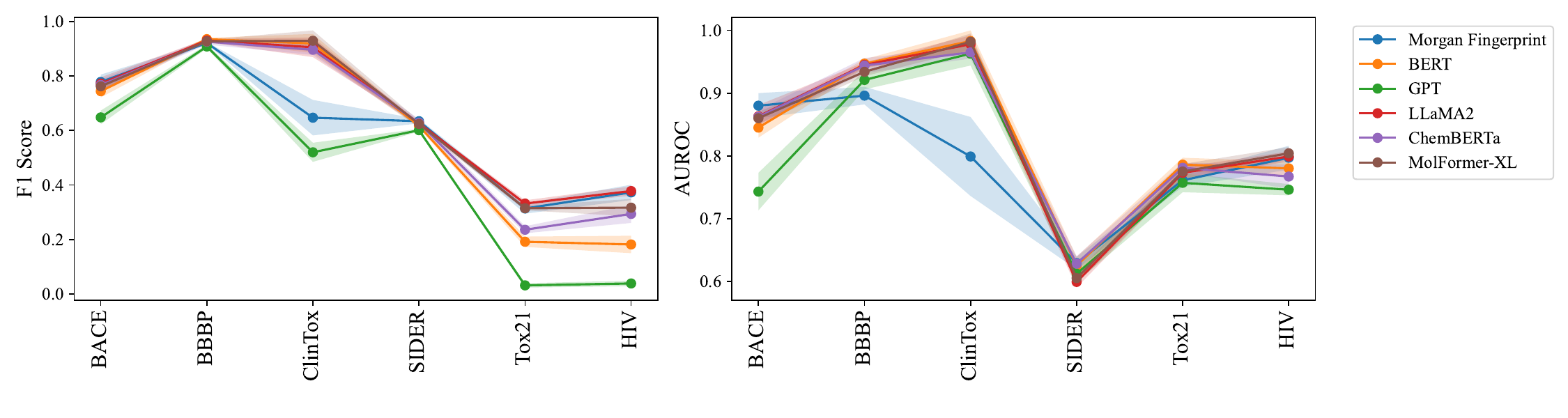} 
        \caption{Classification Task}
        \label{fig:a}
    \end{subfigure}%
    \hfill
    \begin{subfigure}[b]{\textwidth} 
        \centering
        \includegraphics[width=\textwidth]{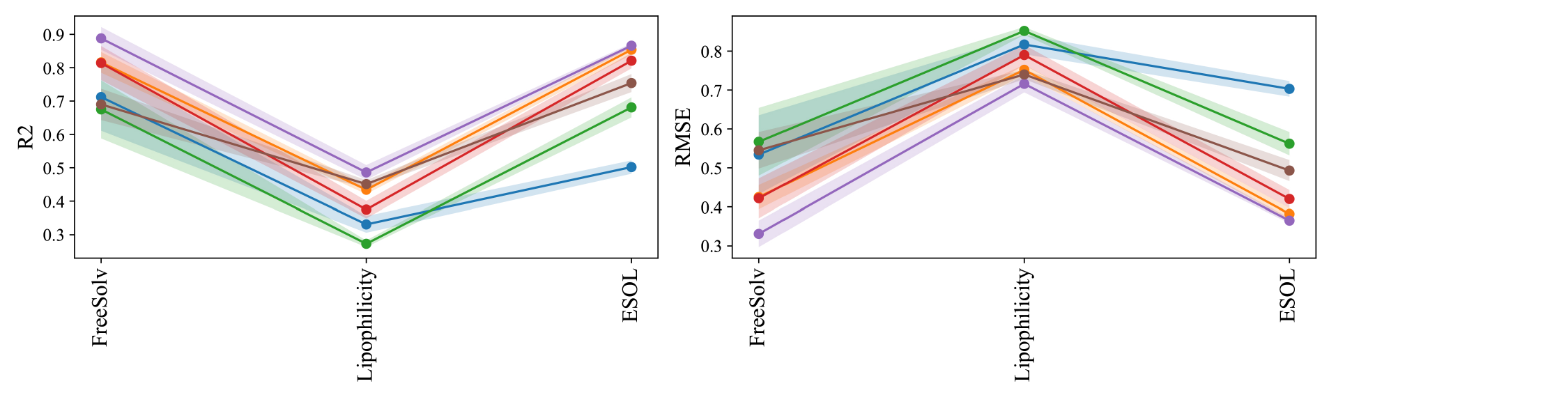} 
        \caption{Regression Task}
        \label{fig:b}
    \end{subfigure}
    \caption{Results on Classification and Regression Tasks. Each Line Represent the Mean Value of 5-Fold Cross Validation While the Shaded Area Shows Their Standard Deviation.}
    \label{fig:2}
\end{figure}

Figure \ref{fig:2}(a), Table \ref{tab:3}, and \ref{tab:4} present our experiments on classification tasks. Surprisingly, LLaMA embeddings achieve comparable performance to established pre-trained models such as MolFormer-XL \cite{ross2022large} and ChemBERTa \cite{chithrananda2020chemberta} across all datasets. Conversely, GPT embeddings underperform in every case. Intriguingly, Morgan FP representations nearly match the performance of other pre-trained methods but are more computationally efficient; generating Morgan FP for a large dataset takes less than a minute without the need for a GPU, whereas LLaMA requires GPUs and processes only 117 molecules per minute (Table \ref{tab:2}). We also tested other classifiers, including SVM and Random Forest, with similar results. The small standard deviation in the evaluation scores indicates that these performance differences are statistically significant. Despite ChemBERTa and MolFormer-XL being pre-trained on millions of compounds from PubChem and ZINC, they perform comparably or, in some instances, less effectively than the BERT model. This showcases the importance of fine-tuning the results of pre-trained models. 
\begin{table*}[!ht]
\centering
\tiny
\caption{Results on Classification Tasks. The Reported Performance Metrics Are the Mean and Standard Deviation of the F1-score and AUROC, Calculated Across the 5-folds. }\label{tab:3}
\begin{tabular}{lcccccc}
\toprule
\textbf{Dataset} 		&\multicolumn{2}{c}{\textbf{BBBP}}	& \multicolumn{2}{c}{\textbf{BACE}}	&\multicolumn{2}{c}{\textbf{HIV}}\\\midrule
\textbf{\# Compounds} 	&	\multicolumn{2}{c}{2039}	&	\multicolumn{2}{c}{1513}	&\multicolumn{2}{c}{41127}	\\
\textbf{Negative:Positive$^*$}	&\multicolumn{2}{c}{$\approx$1:3}	&	\multicolumn{2}{c}{$\approx$1:1}	&\multicolumn{2}{c}{$\approx$28:1}	\\
\midrule
\textbf{Models}				&\textbf{F1-Score}					&\textbf{AUROC}				&\textbf{F1-Score}		 	&\textbf{AUROC}		   			&\textbf{F1-Score}			&\textbf{AUROC}\\
\cmidrule(lr){1-1}\cmidrule(lr){2-3} \cmidrule(lr){4-5} \cmidrule(lr){6-7}
\textbf{Morgan FP}			&0.921 $\pm$ 0.003			&0.896 $\pm$ 0.014			&\textbf{0.778 $\pm$ 0.027}	&\textbf{0.880 $\pm$ 0.020}		&0.373 $\pm$ 0.028			&0.797 $\pm$ 0.019\\
\textbf{BERT}				&\textbf{0.935 $\pm$ 0.005}	&0.947 $\pm$ 0.007			&0.744 $\pm$ 0.023			&0.845 $\pm$ 0.016				&0.182 $\pm$ 0.032			&0.780  $\pm$ 0.011\\
\textbf{ChemBERTa}			&0.926 $\pm$ 0.011			&0.944 $\pm$ 0.012			&0.767 $\pm$ 0.020			&0.862 $\pm$ 0.011				&0.294 $\pm$ 0.033			&0.767 $\pm$ 0.019\\
\textbf{MolFormer-XL}		&0.927 $\pm$ 0.006			&0.934 $\pm$ 0.007			&0.762 $\pm$ 0.012			&0.860 $\pm$ 0.010				&0.317 $\pm$ 0.032			&\textbf{0.804 $\pm$ 0.010}\\
\textbf{GPT}				&0.908 $\pm$ 0.007			&0.921 $\pm$ 0.015			&0.648 $\pm$ 0.025			&0.743 $\pm$ 0.030				&0.039 $\pm$ 0.010			&0.746 $\pm$ 0.009\\
\textbf{LLaMA}				&0.933 $\pm$ 0.006			&\textbf{0.953 $\pm$ 0.009}	&0.766 $\pm$ 0.024			&0.859 $\pm$ 0.017				&\textbf{0.391 $\pm$ 0.013}	&0.802 $\pm$ 0.010\\
\textbf{LLaMA2}				&0.930 $\pm$ 0.006			&0.945 $\pm$ 0.004			&0.772 $\pm$ 0.023			&0.863 $\pm$ 0.018				&0.378 $\pm$ 0.017			&0.799 $\pm$ 0.008\\
\bottomrule
\end{tabular}
\end{table*}

\begin{table*}[!ht]
\centering
\tiny
\caption{Results on Multi-task Classification Tasks. The Reported Performance Metrics Are the Mean and Standard Deviation of the F1-score and AUROC, Calculated Across the 5-folds.}\label{tab:4}
\begin{tabular}{lcccccc}
\toprule
\textbf{Dataset} 		& \multicolumn{2}{c}{\textbf{ClinTox}}	& \multicolumn{2}{c}{\textbf{SIDER}}	&\multicolumn{2}{c}{\textbf{Tox21}}\\\midrule
\textbf{\# Compounds} 	&	\multicolumn{2}{c}{1478}	&	\multicolumn{2}{c}{1427}	&\multicolumn{2}{c}{7831}	\\
\textbf{\# Tasks}	&\multicolumn{2}{c}{2}	&	\multicolumn{2}{c}{27}	&\multicolumn{2}{c}{12}	\\
\midrule
\textbf{Models}				&\textbf{F1-Score}					&\textbf{AUROC}				&\textbf{F1-Score}		 	&\textbf{AUROC}		   			&\textbf{F1-Score}			&\textbf{AUROC}\\
\cmidrule(lr){1-1}\cmidrule(lr){2-3} \cmidrule(lr){4-5} \cmidrule(lr){6-7}
\textbf{Morgan FP}			&0.647 $\pm$ 0.065			&0.799 $\pm$ 0.063			&\textbf{0.634 $\pm$ 0.008}	&\textbf{0.629 $\pm$ 0.01}		&0.314 $\pm$ 0.019			&0.761 $\pm$ 0.010\\
\textbf{BERT}				&0.919 $\pm$ 0.035			&0.983 $\pm$ 0.017			&0.617 $\pm$ 0.008			&0.625 $\pm$ 0.014				&0.192 $\pm$ 0.019			&\textbf{0.786 $\pm$ 0.011}\\
\textbf{ChemBERTa}			&0.896 $\pm$ 0.019			&0.965 $\pm$ 0.01			&0.628 $\pm$ 0.014			&0.628 $\pm$ 0.012				&0.236 $\pm$ 0.013			&0.781 $\pm$ 0.008\\
\textbf{MolFormer-XL}		&\textbf{0.929 $\pm$ 0.038}	&\textbf{0.982 $\pm$ 0.013}	&0.624 $\pm$ 0.012			&0.605 $\pm$ 0.009				&0.315 $\pm$ 0.008			&0.775 $\pm$ 0.012\\
\textbf{GPT}				&0.520 $\pm$ 0.035			&0.963 $\pm$ 0.019			&0.601 $\pm$ 0.005			&0.612 $\pm$ 0.013				&0.032 $\pm$ 0.008			&0.757 $\pm$ 0.015\\
\textbf{LLaMA}				&0.881 $\pm$ 0.053			&0.980 $\pm$ 0.008			&0.627 $\pm$ 0.007			&0.605 $\pm$ 0.008				&\textbf{0.339 $\pm$ 0.015}	&0.774 $\pm$ 0.010\\
\textbf{LLaMA2}				&0.905 $\pm$ 0.036			&0.978 $\pm$ 0.014			&0.627 $\pm$ 0.004			&0.599 $\pm$ 0.009				&0.332 $\pm$ 0.012			&0.773 $\pm$ 0.009\\
\bottomrule
\end{tabular}
\end{table*}

\subsubsection{Results on Regression Tasks}
Figure \ref{fig:2}(a) and Table \ref{tab:5} present the evaluation results for the regression tasks. Similar to the classification results, GPT underperforms relative to other models, and in some instances, it even falls short of Morgan Fingerprint's performance. ChemBERTa consistently emerges as the top-performing model for regression across all tested datasets. BERT and LLaMA exhibit performances that are closely comparable to ChemBERTa in the regression tasks. Additionally, we observed a general decline in the performance of all methods when applied to larger datasets, such as Lipophilicity.

\begin{table*}[!ht]
\centering
\tiny
\caption{Results on Regression Tasks. The Reported Performance Metrics Are the Mean and Standard Deviation of the RMSE and R$^2$, Calculated Across the 5-folds.}\label{tab:5}
\begin{tabular}{lcccccc}
\toprule
\textbf{Dataset} 		& \multicolumn{2}{c}{\textbf{FreeSolv}}	& \multicolumn{2}{c}{\textbf{Lipophilicity}}	&\multicolumn{2}{c}{\textbf{ESOL}}\\\midrule
\textbf{\# Compounds} 	&	\multicolumn{2}{c}{642}	&	\multicolumn{2}{c}{4200}	&\multicolumn{2}{c}{1128}	\\
\midrule
\textbf{Models}				&\textbf{RMSE}					&\textbf{R$^2$}				&\textbf{RMSE}		 	&\textbf{R$^2$}		   			&\textbf{RMSE}			&\textbf{R$^2$}\\
\cmidrule(lr){1-1}\cmidrule(lr){2-3} \cmidrule(lr){4-5} \cmidrule(lr){6-7}
\textbf{Morgan FP}			&0.534 $\pm$ 0.101			&0.712 $\pm$ 0.101			&0.817 $\pm$ 0.025			&0.331 $\pm$ 0.025				&0.703 $\pm$ 0.020			&0.502 $\pm$ 0.020\\
\textbf{BERT}				&0.425 $\pm$ 0.031			&0.816 $\pm$ 0.031			&0.752 $\pm$ 0.013			&0.434 $\pm$ 0.013				&0.382 $\pm$ 0.015			&0.854 $\pm$ 0.015\\
\textbf{ChemBERTa}			&\textbf{0.331 $\pm$ 0.034}	&\textbf{0.888 $\pm$ 0.034}	&\textbf{0.716 $\pm$ 0.022}	&\textbf{0.486 $\pm$ 0.022}		&\textbf{0.365 $\pm$ 0.007}	&\textbf{0.866 $\pm$ 0.007}\\
\textbf{MolFormer-XL}		&0.545 $\pm$ 0.047			&0.690 $\pm$ 0.047			&0.740 $\pm$ 0.012			&0.451 $\pm$ 0.012				&0.493 $\pm$ 0.027			&0.754 $\pm$ 0.027\\
\textbf{GPT}				&0.567 $\pm$ 0.087			&0.675 $\pm$ 0.087			&0.852 $\pm$ 0.010			&0.273 $\pm$ 0.010				&0.562 $\pm$ 0.030			&0.681 $\pm$ 0.030\\
\textbf{LLaMA}				&0.483 $\pm$ 0.036			&0.758 $\pm$ 0.036			&0.785 $\pm$ 0.015			&0.382 $\pm$ 0.015				&0.425 $\pm$ 0.013			&0.818 $\pm$ 0.013\\
\textbf{LLaMA2}				&0.422 $\pm$ 0.051			&0.814 $\pm$ 0.051			&0.790 $\pm$ 0.026			&0.375 $\pm$ 0.026				&0.420 $\pm$ 0.023			&0.821 $\pm$ 0.023\\
\bottomrule
\end{tabular}
\end{table*}

\subsubsection{Results on Link Prediction Tasks}
Table \ref{tab:6} presents the results for the link prediction tasks on DDI networks. LLaMA consistently outperforms all other models across both datasets by a significant margin. Notably, Morgan FP surpasses the performance of embeddings from pre-trained models. It appears that the size of the embeddings impacts model performance, as larger embeddings generally yield better results. Nevertheless, despite having the same size, there are still noticeable performance differences between the LLaMA and LLaMA2 models.

\begin{table*}[!ht]
\centering
\tiny
\caption{Results on Link Prediction Tasks. The Reported Performance Metrics Are the Mean and Standard Deviation of the AUROC and AUPR, Calculated Across the 10 Runs.}\label{tab:6}
\begin{tabular}{lccccc}
\toprule
\textbf{Dataset} 		& \multicolumn{2}{c}{\textbf{BioSnap}}	& \multicolumn{2}{c}{\textbf{DrugBank}}	\\\midrule
\textbf{\# Nodes} 	&	\multicolumn{2}{c}{1320}	&	\multicolumn{2}{c}{1690}		\\
\textbf{\# Edges} 	&	\multicolumn{2}{c}{41577}	&	\multicolumn{2}{c}{190609}		\\
\textbf{Average node degree} 	&	\multicolumn{2}{c}{64.087}	&	\multicolumn{2}{c}{224.38}		\\

\midrule
\textbf{Models}				&\textbf{AUROC}					&\textbf{AUPR}				&\textbf{AUROC}				&\textbf{AUPR}		   			\\
\cmidrule(lr){1-1} \cmidrule(lr){2-3}\cmidrule(lr){4-5} 
\textbf{Morgan FP}			&0.871 $\pm$ 0.00				&0.847 $\pm$ 0.00			&0.876 $\pm$ 0.00			&0.855 $\pm$ 0.00\\
\textbf{BERT}				&0.621 $\pm$ 0.02				&0.563 $\pm$ 0.08			&0.660 $\pm$ 0.02			&0.639 $\pm$ 0.01\\
\textbf{ChemBERTa}			&0.527 $\pm$ 0.02				&0.547 $\pm$ 0.08			&0.519 $\pm$ 0.02			&0.457 $\pm$ 0.01\\
\textbf{MolFormer-XL}		&0.550 $\pm$ 0.02				&0.701 $\pm$ 0.08			&0.611 $\pm$ 0.02			&0.644 $\pm$ 0.01\\
\textbf{GPT}				&0.856 $\pm$ 0.06				&0.812 $\pm$ 0.08			&0.836 $\pm$ 0.05			&0.748 $\pm$ 0.09\\
\textbf{LLaMA}				&0.921 $\pm$ 0.00				&0.884 $\pm$ 0.02			&0.927 $\pm$ 0.00			&0.872 $\pm$ 0.01\\
\textbf{LLaMA2}				&\textbf{0.941 $\pm$ 0.00}		&\textbf{0.902 $\pm$ 0.02}	&\textbf{0.961 $\pm$ 0.00}	&\textbf{0.933 $\pm$ 0.01}\\
\bottomrule
\end{tabular}
\end{table*}

\subsubsection{Ablation Study}

\begin{figure}[H]
    \centering
    \begin{subfigure}[b]{\textwidth} 
        \centering
        \includegraphics[width=\textwidth]{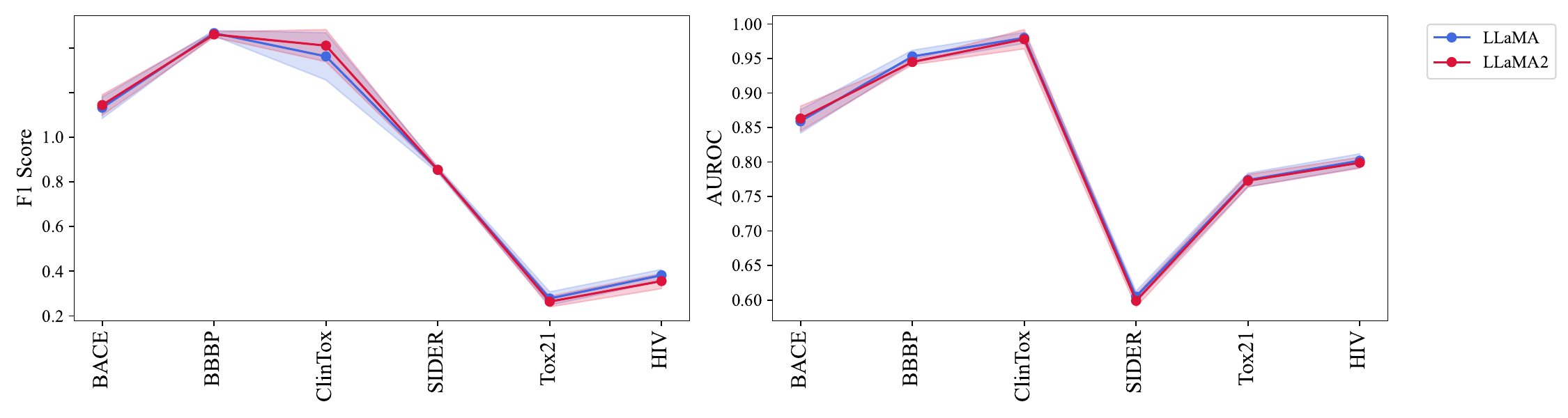} 
        \caption{Classification Tasks}
    \end{subfigure}%
    \hfill
    \begin{subfigure}[b]{\textwidth} 
        \centering
        \includegraphics[width=\textwidth]{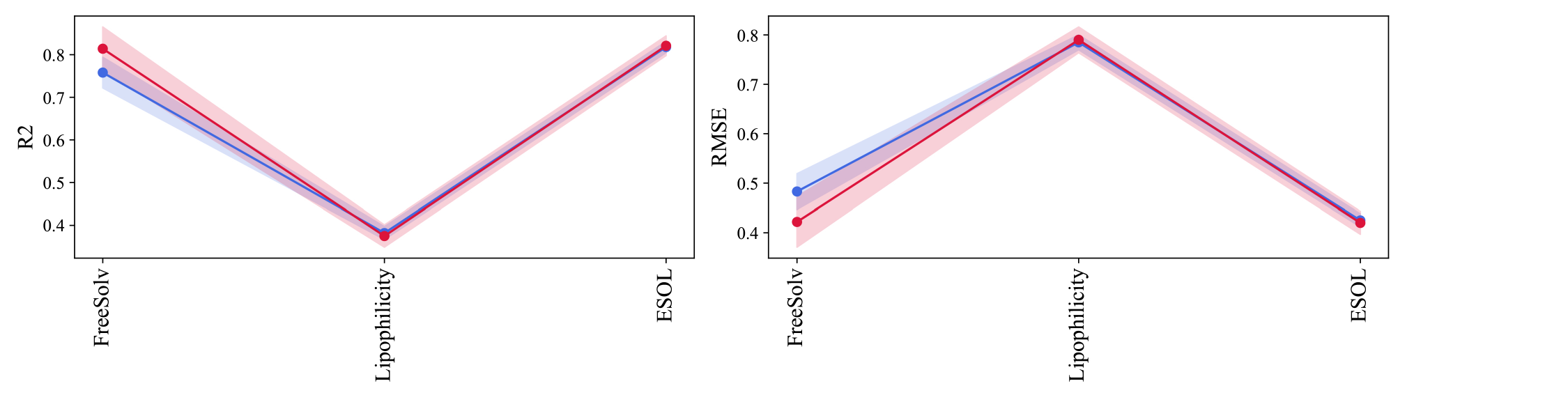} 
        \caption{Regression Tasks}
    \end{subfigure}
    \caption{Comparison of LLaMA and LLaMA2 Performance }
    \label{fig:3}
\end{figure}

\textbf{LLaMA Vs LLaMA2}
Figure \ref{fig:3} compares the LLaMA and LLaMA2 models. The performance of these two models is similar, mainly across various tasks. However, there are notable differences in specific instances. For example, in the link prediction tasks (Table \ref{tab:6}), LLaMA2 outperforms LLaMA. This trend is also observed in classification and regression tasks, where LLaMA2 generally matches or exceeds the performance of LLaMA. Both models share similar architecture and training presets. Nevertheless, LLaMA2 has been trained on 40\% more data and supports twice the context length of its predecessor, enhancing its capability to understand more complex language structures \cite{touvron2023llama, touvron2023llama2}. 

\begin{figure}[H]
    \centering
    \begin{subfigure}[b]{\textwidth} 
        \centering
        \includegraphics[width=\textwidth]{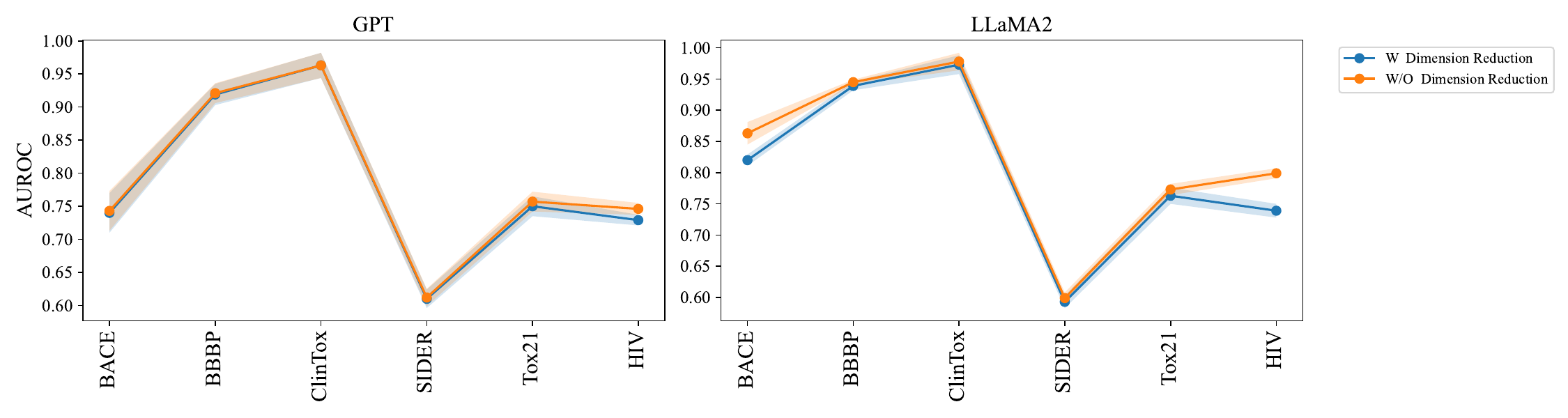} 
        \caption{Classification Tasks}
    \end{subfigure}%
    \hfill
    \begin{subfigure}[b]{\textwidth} 
        \centering
        \includegraphics[width=\textwidth]{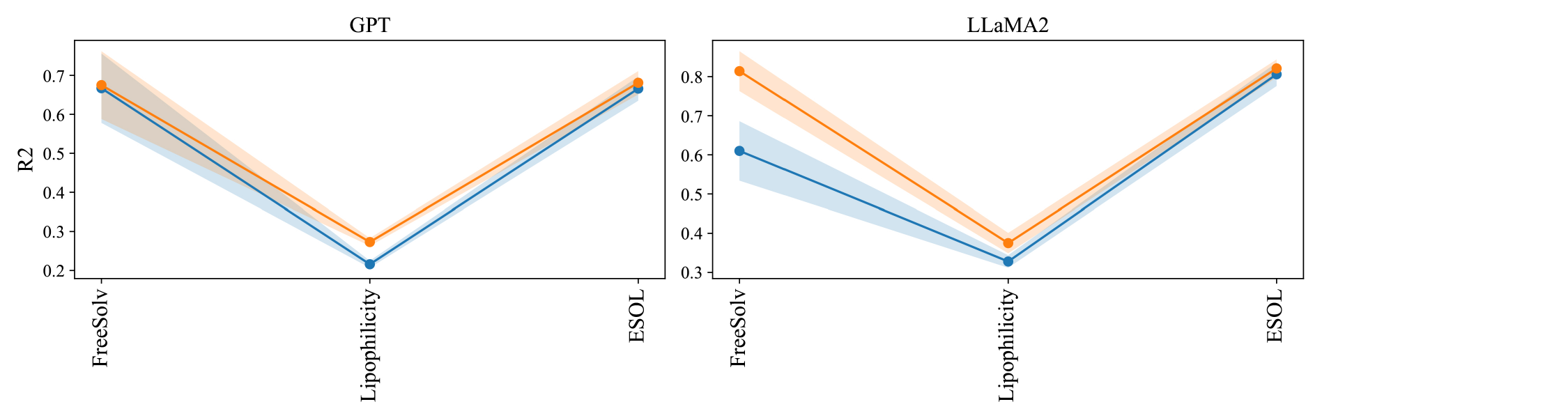} 
        \caption{Regression Tasks}
    \end{subfigure}
    \caption{Effect of Dimension Reduction on The Performance of LLMs}
    \label{fig:4}
\end{figure}

\textbf{Dimension Reduction}
We investigated the impact of dimension reduction on LLMs with substantial embedding sizes, as illustrated in Figure \ref{fig:4}. Using Principal Component Analysis (PCA) for dimension reduction, we experimented with various reduction sizes. Our findings indicate that the impact of dimension reduction on the classification performance of GPT and LLaMA models is minimal, although there is a noticeable decrease in performance post-reduction.
In contrast, for regression tasks, dimension reduction significantly lowers the performance of the models. This suggests a correlation between the size of the embeddings in LLMs and their effectiveness in handling regression tasks. 

\begin{figure}[ht]
    \centering
    \includegraphics[width=\textwidth]{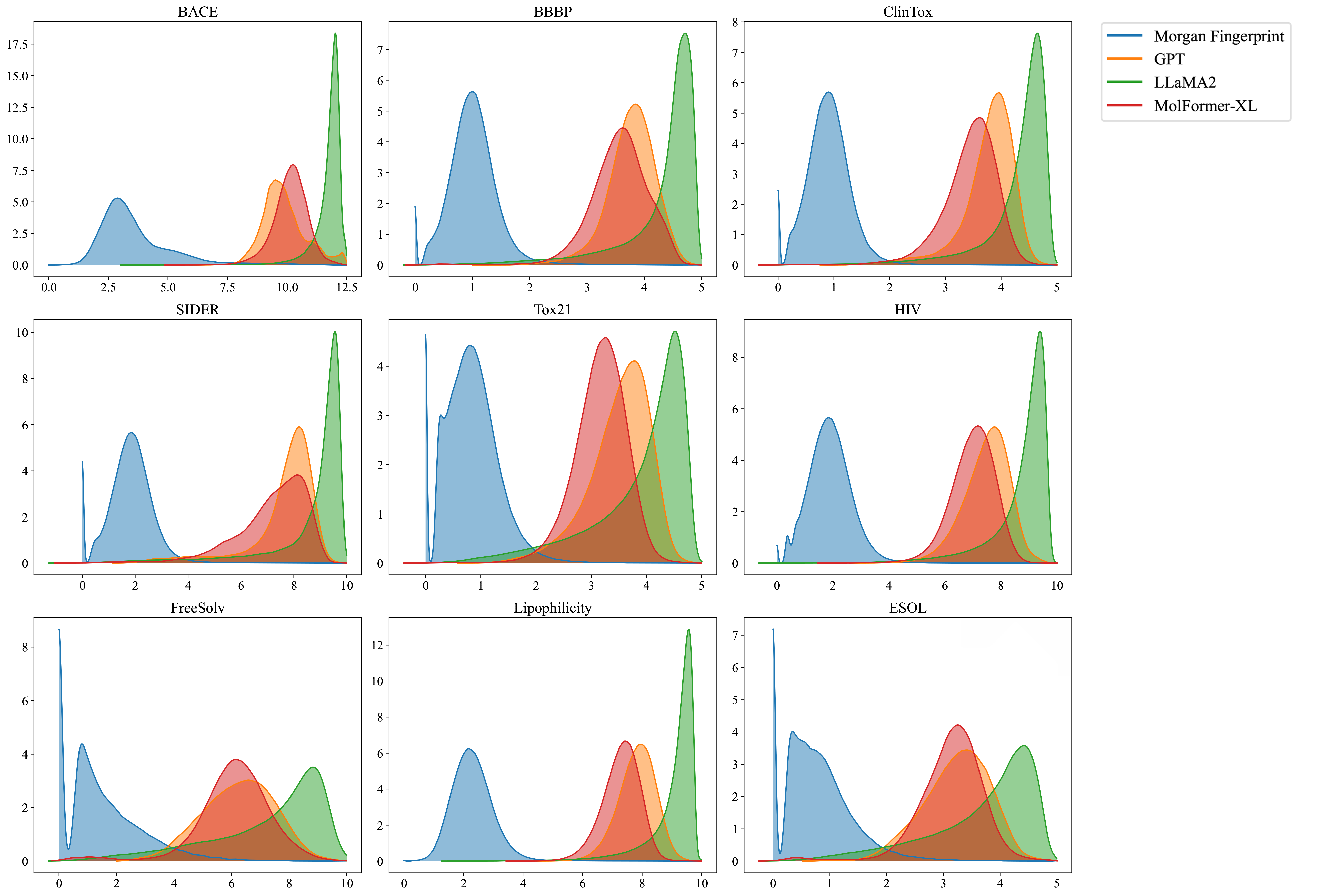}
    \caption{Anisotropy problem of LLM Models Embedding}
    \label{fig:5}
\end{figure}

\begin{figure}[H]
    \centering
    \includegraphics[width=\textwidth]{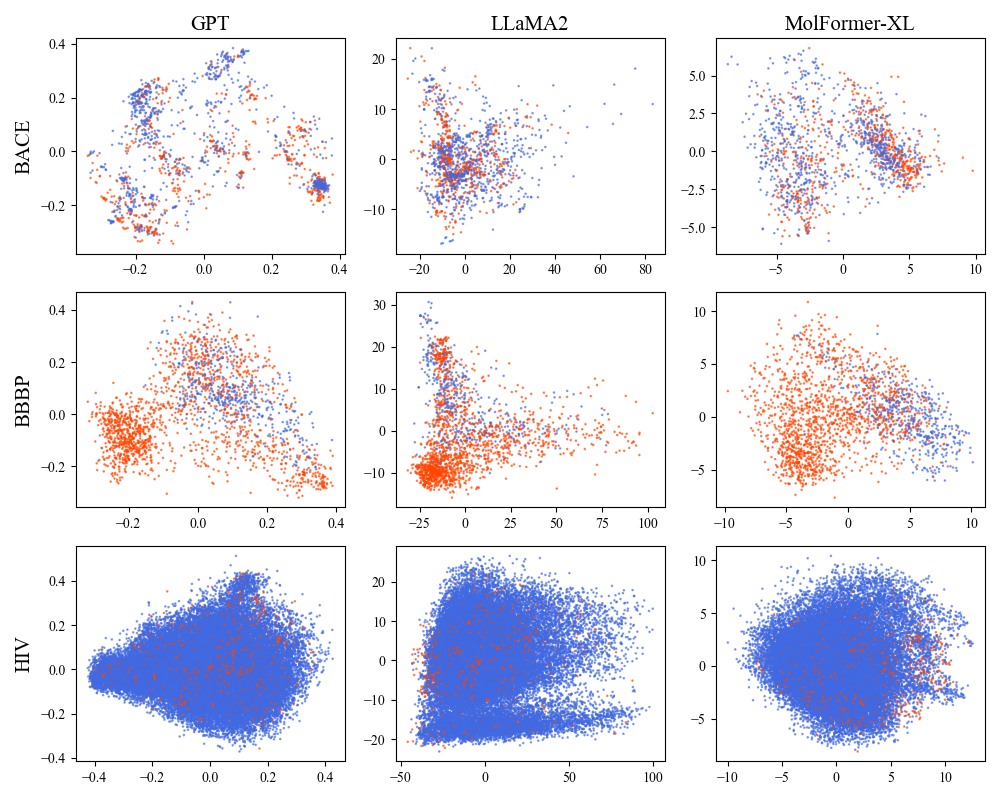}
    \caption{PCA Representation Embedding for Classification Task. Red Represent Positive Samples While Blue Represent Negative Samples}
    \label{fig:6}
\end{figure}

\textbf{LLM And Anistropy}
It is well documented that LLM embeddings have the isotropy problem \cite{timkey2021all,kovaleva2021bert,rudman2021isoscore}. Our comparative analysis also reveals that LLMs embeddings demonstrate a higher degree of anisotropy than pre-trained embeddings and Morgan FP (Figure \ref{fig:5}). This is evident since the distribution of cosine similarity of embeddings is more closely grouped together in their representation  (Figure \ref{fig:5}). However, our experiments indicate that better isotropy does not imply a performance gain in machine-learning tasks. As can be seen, the cosine similarity distribution of LLaMA2 embeddings is a lot narrower than GPT and Morgan FP; however, LLaMA2 outperforms both models in most cases.
As illustrated in Figure \ref{fig:6}, we also noticed that the PCA representation of GPT's embeddings is predominantly concentrated within a range smaller than 1. This observation also suggests a high likelihood that the GPT embeddings have been pre-normalized.

\begin{figure}[H]
    \centering
    \begin{subfigure}[b]{\textwidth} 
        \centering
        \includegraphics[width=\textwidth]{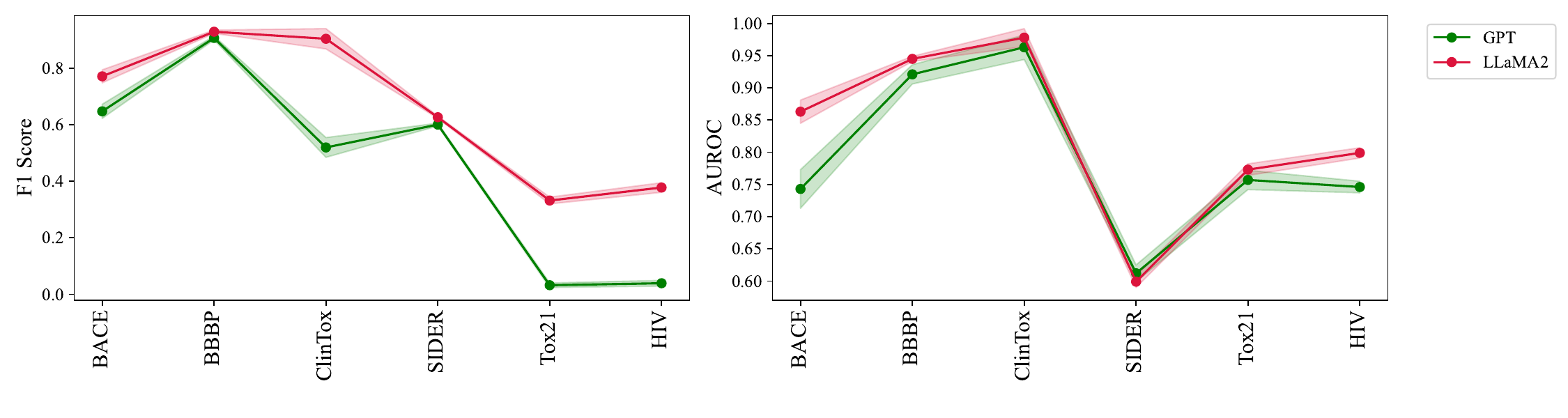} 
        \caption{Classification Tasks}
    \end{subfigure}%
    \hfill
    \begin{subfigure}[b]{\textwidth} 
        \centering
        \includegraphics[width=\textwidth]{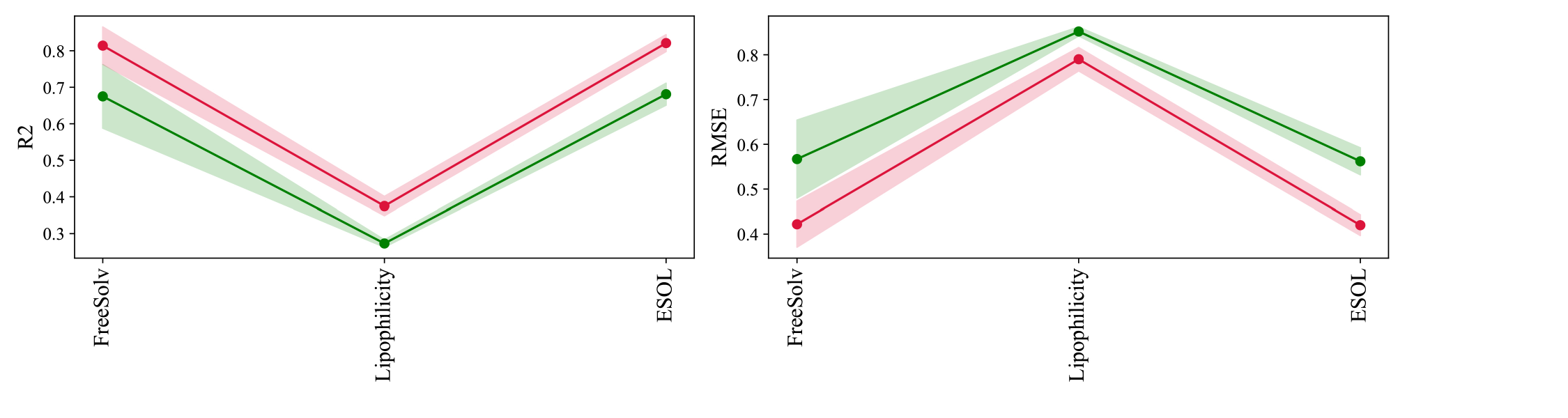} 
        \caption{Regression Tasks}
    \end{subfigure}
    \caption{Comparison of LLaMA2 and GPT. }
    \label{fig:7}
\end{figure}

\textbf{GPT Vs LLaMA}
Figure \ref{fig:7} demonstrates that LLaMA consistently outperforms GPT across all datasets by a significant margin. This raises the question of whether these differences are due to the architectural design or the specific training of the models. As outlined in the GPT-4 technical report, GPT models are capable of interpreting SMILES strings. Notably, approximately 2.5\% of the LLaMA training dataset, as reported in \cite{touvron2023llama, zhao2023survey}, consists of scientific material primarily sourced from arXiv, including bioinformatics papers. 

Both LLaMA and GPT models utilize a transformer-based architecture with a heavy reliance on self-attention mechanisms and a decoder-only configuration. However, the opaque nature of GPT as a "black box" model complicates direct comparisons with LLaMA regarding whether their efficiency stems solely from architecture or pre-training specifics. Nonetheless, considering their training on SMILES strings, the data from Figure \ref{fig:7} and Table \ref{tab:6} suggest that the LLaMA architecture is particularly adept at handling complex language structures like SMILES strings. Furthermore, Table \ref{tab:1} reveals that while the LLaMA2 tokenizer may not perform as well as the MolFormer tokenizer, it tokenizes SMILES strings more effectively than BERT. Unfortunately, we cannot compare the GPT tokenizer directly with other models due to limitations in OpenAI's API access.

\begin{figure}[H]
    \centering
    \includegraphics[width=\textwidth]{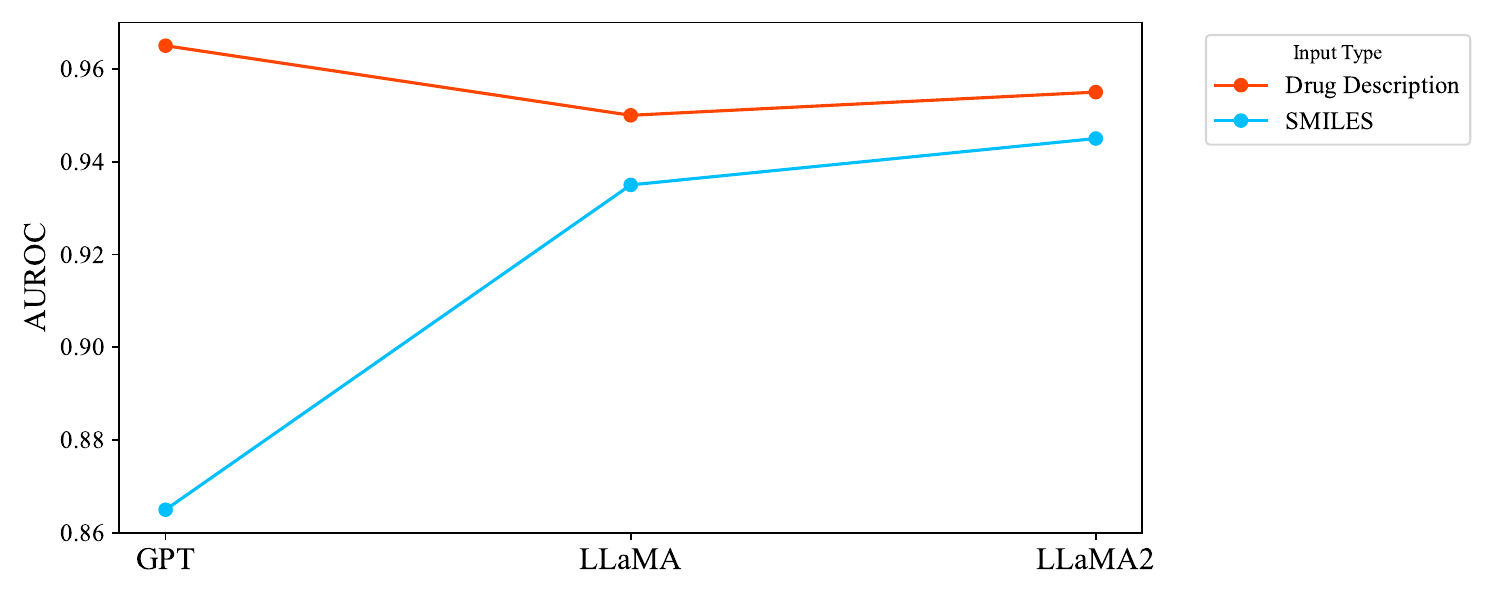}
    \caption{Impact of Drug Description for DDI Prediction on BioSnap Dataset}
    \label{fig:8}
\end{figure}

\textbf{Link Prediction with SMILES VS Drug Description}
We also extracted the text-format drug description information of drugs from the DrugBank database. Drug description embedding in DDI prediction significantly outperforms using SMILES strings when leveraging LLMs. This enhancement is consistent with applying LLMs pre-trained on general text data, as depicted in Figure \ref{fig:8}. When applied to drug descriptions closer to natural language, GPT outperforms the LLaMA models on both datasets and uses both AUROC and AUPRC metrics.

\section{Conclusions}

In summary, this research underscores the potential of LLMs like GPT and LLaMA for molecular embedding. We specifically recommend LLaMA models over GPT due to their superior performance in generating molecular embeddings from SMILES strings, which is notable in our studies. These findings suggest that LLaMA could be particularly effective in predicting molecular properties and drug interactions. Although models like LLaMA and GPT are not explicitly designed for SMILES string embedding—unlike specialized models such as ChemBERTa and MolFormer-XL—they still demonstrate competitive performance. Our work lays the groundwork for future improvements in utilizing LLMs for molecular embedding. Future efforts will focus on enhancing the quality of molecular embeddings derived from LLMs inspired by natural language sentence embedding techniques, such as fine-tuning and modifications to LLaMA tokenization.

\section*{Acknowledgements}
Not applicable.

\bibliography{Main}

\end{document}